\documentclass[useAMS,usenatbib]{mn2e}

\usepackage{psfig}
\usepackage{epsfig}
\usepackage[varg]{txfonts}
\def\ltsima{$\; \buildrel < \over \sim \;$}
\def\lsim{\lower.5ex\hbox{\ltsima}}
\def\gtsima{$\; \buildrel > \over \sim \;$}
\def\gsim{\lower.5ex\hbox{\gtsima}}

\begin{document}

\title[Irradiated self-gravitating discs]{Stability of self-gravitating discs under irradiation}
\author[Rice et al.] {W.~K.~M. Rice$^{1}$\thanks{E-mail: wkmr@roe.ac.uk}, P.~J. Armitage$^{2,3}$,
G.R. Mamatsashvili$^{1,4}$, 
G. Lodato$^{5}$ and C.~J. Clarke$^{6}$\\
$^1$Scottish Universities Physics Alliance (SUPA), Institute for Astronomy, University of Edinburgh, Blackford Hill, Edinburgh EH9 3HJ \\
$^2$JILA, 440 UCB, University of Colorado, Boulder, CO 80309-0440, USA \\
$^3$Department of Astrophysical and Planetary Sciences, University of Colorado, Boulder, USA \\
$^4$Georgian National Astrophysical Observatory, Il. Chavchavadze State University, 2a Kazbegi Ave., Tbilisi 0160, Georgia \\
$^5$Universit\`a degli Studi di Milano, Dipartimento di Fisica, via Celoria 16, I-20133 Milano, Italy \\
$^6$Institute of Astronomy, Madingley Road, Cambridge, CB3 0HA, UK\\}

\maketitle
  
\begin{abstract}
Self-gravity becomes competitive as an angular momentum transport process in accretion 
discs at large radii, where the temperature is low enough that external irradiation 
likely contributes to the thermal balance. Irradiation is known to weaken the strength of disc self-gravity, 
and can suppress it entirely if the disc is maintained above the threshold for linear 
instability. However, its impact on the susceptibility of the disc to fragmentation is less clear. We 
use two-dimensional numerical simulations to investigate the evolution of self-gravitating discs 
as a function of the local cooling time and strength of irradiation. In the 
regime where the disc does not fragment, we show that local thermal equilibrium 
continues to determine the stress - which can be represented as an effective viscous $\alpha$ - out to very long cooling times, $\tau_c = 240 \Omega^{-1}$.
In this regime, it is also found that the power spectrum of the perturbations is
uniquely set by this effective viscous $\alpha$ and not by the cooling rate. 
Fragmentation occurs for $\tau_c < \beta_{\rm crit} \Omega^{-1}$, where $\beta_{\rm crit}$ is 
a weak function of the level of irradiation. We find that $\beta_{\rm crit}$ declines 
by approximately a factor of two, as irradiation is increased from zero 
up to the level where instability is almost quenched. The numerical results 
imply that irradiation cannot generally avert fragmentation of self-gravitating discs 
at large radii; if other angular momentum transport sources are weak mass will 
build up until self-gravity sets in, and fragmentation will ensue.  
\end{abstract}


\section{Introduction}
Self-gravity may be important in the cool outer regions of both 
protostellar and Active Galactic Nuclei (AGN) accretion discs, 
where the combined effects of pressure and shear cannot stabilize 
the flow against gravitational instability. The local linear stability 
of self-gravitating discs is simple \citep{toomre64}: it depends upon 
the parameter,
\begin{equation}
 Q \equiv \frac{c_s \kappa}{\pi G \Sigma},
\end{equation}
where $c_s$ and $\Sigma$ are the sound speed and surface density of 
a low-mass disc ($M_{\rm disc} \ll M_*$), and $\kappa$ is the epicyclic frequency
which, in a Keplerian disc, is the same as the angular 
velocity $\Omega$. The non-linear behaviour, however, is complex. 
A self-gravitating disc can either fragment into bound objects, or 
attain a ``stable" self-gravitating state in which angular momentum 
transport results in accretion and energy dissipation. Determining 
what sets the boundary between these outcomes is important for 
understanding angular momentum transport \citep{armitage11} and planet formation in 
protoplanetary disks \citep{boss97}, and star formation in discs around AGN \citep{bonnell08}.

Since gravity is a long-range force  
angular momentum transport via self-gravity could in principle be a non-local 
process \citep{balbus99}. Global numerical simulations \citep{lodato04,cossins09}, 
however, show that disc self-gravity can be approximated surprisingly 
well using a model in which turbulent heating associated with 
angular momentum transport is {\em locally} balanced by radiative losses. 
The use of this local approximation \citep{paczynski78,lin87} greatly 
simplifies the description of the disc. In a local description, thermal equilibrium 
implies that the stress, 
measured via the equivalent Shakura-Sunyaev $\alpha$ parameter 
\citep{shakura73}, is inversely proportional to the 
time scale on which the disc can radiate its thermal energy \citep{gammie01}. 
The stability of the disc against fragmentation also depends 
upon its cooling time \citep{shlosman87}. Writing $t_{\rm cool} = \beta \Omega^{-1}$, 
two-dimensional local numerical simulations show that fragmentation of a self-luminous 
disc (one in which the only source of heat is furnished by accretion) 
ensues whenever $\beta < \beta_{\rm crit} \simeq 3$ \citep[][assuming a 2D adiabatic index 
$\gamma = 2$]{gammie01}. This minimum cooling time before fragmentation is equivalent to a maximum 
$\alpha_{\rm crit} \simeq 0.1$ that a stable self-gravitating disc can sustain \citep{rice05}. 
Consistent results for the location of the fragmentation boundary are obtained from two-dimensional global 
simulations, provided that care is taken to avoid prompt fragmentation during the initial 
growth of instability \citep{paardekooper11}. In three-dimensional global simulations 
the situation is less clear. There is evidence that such discs are somewhat 
less stable against fragmentation than their two-dimensional counterparts \citep{meru11}, 
although a critical cooling time does nonetheless appear to exist \citep{lodato11,rice11}.
We also note that strong temperature dependence of the opacity 
can modify the fragmentation boundary substantially \citep{johnson03,cossins09b}. 
This work deals exclusively with simpler temperature-independent disc cooling. 

Applying these results for self-luminous discs to protostellar and AGN discs, 
fragmentation is predicted to occur for $r \gsim 70-100~{\rm au}$ 
in protostellar discs \citep{matzner05,rafikov05,stamatellos07,boley09,clarke09,
rafikov09,rice10}, and at $r \gsim 0.06 \ (M_{\rm BH} / 10^6 \ M_\odot)^{1/3} \ {\rm pc}$ 
for AGN \citep{rafikov09,goodman03,levin07}. At these large radii, however, the 
temperature of a self-luminous disc is small enough -- about 10~K -- that 
the thermal balance of a real disc in a star forming region or galactic 
nucleus may often be influenced by external irradiation (or, in the AGN case, 
by internal heating due to embedded stars). Here, we use numerical
simulations to revisit 
the stability of self-gravitating discs, taking into account 
that irradiation as well as dissipation of accretion energy 
may contribute to the thermal balance. Our goal is to determine 
whether, in an irradiated disc, the fragmentation boundary is determined 
by a minimum cooling time or by a maximum $\alpha$. These quantities, which 
are no longer equivalent in a disc subject to irradiation, are both physically  
important: $\beta$ measures the time scale on which overdense 
clumps can radiate thermal energy and collapse further, while $\alpha$ 
is related to the amplitude and non-linearity of density perturbations. 
We also address the question of whether an annulus of the disc, 
initially assumed to be stabilized against self-gravity by irradiation, 
is able to viscously respond to an increase in surface density without 
entering the fragmentation regime. This work expands on earlier work
by \citet{cai08} in which irradiation from an envelope around a protoplanetary disc
was shown to weaken the gravitational instability, and
is complementary to the analytic discussion in \citet{kratter11}.

The plan of this paper is as follows. In Section~2 we describe the 
local (shearing-sheet) simulations that we perform to investigate
the evolution of irradiated self-gravitating discs, in Section~3 we
discuss the results of these simulations, and in Section~4 we
discuss some implications of our results.

\section{Method}
To investigate the evolution of self-gravitating discs in the presence of 
external irradiation we use the Pencil Code, a 
finite difference code that uses sixth-order centered spatial
derivatives and a third-order Runge-Kutta time-stepping scheme
(see \citet{brandenburg03} for details). As in \citet{gammie01},
we use a ``shearing sheet", or ``local", model in which the disc dynamics
is studied in a local Cartesian coordinate frame 
corotating with the angular velocity, $\Omega$, of the disc at some radius from
the central star. In this coordinate
frame, the unperturbed differential rotation of the disc manifests itself
as a parallel azimuthal flow with a constant velocity shear in
the radial direction. We assume Keplerian rotation and hence use a shear parameter
of $q = 1.5$ (such that the $y$-component of the fluid velocity is $u_y = - q \Omega x$).  The unperturbed background surface density, $\Sigma_o$,
and two-dimensional pressure, $P_o$, are assumed to be spatially constant. A 
Coriolis force is included to take into account the effects of the 
coordinate frame rotation.

To use the ``local" model we need to assume that the disc is cool and hence 
thin ($H/r \simeq c_s/(\Omega r) \ll 1$) and, as in \citet{gammie01},
we also assume that it is razor-thin. The continuity equation and 
equations of motion in this model are
\begin{eqnarray}
\frac{\partial \Sigma}{\partial t} + \nabla \cdot(\Sigma \textbf {\em u}) -
q \Omega x \frac{\partial \Sigma}{\partial y} &=& 0 \nonumber \\
\frac{\partial u_x}{\partial t} + (\textbf {\em u} \cdot \nabla)u_x - q \Omega x 
\frac{\partial u_x}{\partial y} &=& - \frac{1}{\Sigma}\frac{\partial P}{\partial x} +
2 \Omega u_y - \frac{\partial \phi}{\partial x} + f_{\nu x} \\
\frac{\partial u_y}{\partial t} + (\textbf {\em u} \cdot \nabla)u_y - q \Omega x
\frac{\partial u_y}{\partial y} &=& - \frac{1}{\Sigma}\frac{\partial P}{\partial y} +
(q - 2) \Omega u_x - \frac{\partial \phi}{\partial y} + f_{\nu y} \nonumber
\label{eq:eqn_motion}
\end{eqnarray} 
where $\textbf{\em u}(u_x, u_y)$ is velocity relative to the background parallel
shear flow $\textbf {\em u}_o(0, -q \Omega x)$, 
$P$ is the two-dimensional pressure,
$\Sigma$ is the surface density, $\phi$ is the gravitational potential of the gas sheet,
and $\textbf{\em f}_\nu(f_{\nu x}, f_{\nu y})$ is a viscosity term that includes both a 
kinematic viscosity and a bulk viscosity for resolving shocks. Since this disc is 
razor-thin, Poisson's equation becomes
\begin{equation}
\nabla^2 \phi = 4 \pi G \Sigma \delta(z).
\label{eq:Poisson}
\end{equation}
This can be solved by Fourier transfoming (using the standard 
Fast Fourier Transform (FFT) technique) the surface density from the 
($x$, $y$) plane to the ($k_x$, $k_y$) plane giving
\begin{equation}
\phi(k_x,k_y,t) = -\frac{2 \pi G \Sigma(k_x,k_y,t)}{k},
\label{eq:fourier}
\end{equation}
where $\phi(k_x,k_y,t)$ and $\Sigma(k_x,k_y,t)$ are the Fourier
transforms of the gravitational potential and surface density,
and $k = \sqrt{k_x^2 + k_y^2}$.  It should be noted that 
in the shearing sheet approximation, the radial 
wavenumber, $k_x$, is no longer constant, but varies with time as
$k_x(t) = k_x(0) + q \Omega k_y t$.  For a two-dimensional
domain of size $L \times L$ with resolution $N \times N$, the only
Fourier harmonics that will be present at time $t$ will have wavenumbers $k_y = 2 \pi n_y/L$
and $k_x = 2 \pi n_x/L + q \Omega t'(2 \pi n_y/L)$, where $t' = 
mod[t,1/(q \Omega | n_y |)]$, and integer numbers $n_x$ and $n_y$ lie
in the range $-N/2 \le n_x, n_y \le N/2$. As in \citet{gammie01}
we only use wavenumbers that satisfy $k < \pi N/(L \sqrt{2})$
which is the largest circular region in Fourier space that is always
available, and ensures that the gravitational force is isotropic on
small scales. 

The equation of state is essentially the standard
\begin{equation}
P = (\gamma - 1)U,
\label{eq:eos}
\end{equation}
where $U$ is the two-dimensional internal energy per unit volume, and $\gamma$ is the 
two-dimensional adiabatic index.  We use $\gamma = 1.6$ which
can be mapped to a three-dimensional
adiabatic index of between $\sim 1.6$ and $1.9$ depending on whether the disc is strongly
self-gravitating or not (e.g., \citet{gammie01}). For protostellar 
discs, a smaller value of $\gamma$ may be more appropriate.  However, since the work here is
not restricted only to protostellar discs, we have chosen to use $\gamma = 1.6$. 
Rather than solving for internal energy, the 
Pencil Code actually solves for the specific entropy, $s$ (see \citet{brandenburg03} for details).  
We use a cooling function similar to that used by \citet{gammie01}
but modified by the inclusion of a heating term that represents 
external irradiation.  When written in terms of specific entropy,
the combined heating and cooling function that is imposed is
\begin{equation}
\Sigma T \frac{\partial s}{\partial t} = - \frac{\Sigma \left(c_s^2 - c_{so}^2\right)}{\gamma 
(\gamma - 1) \tau_c},
\label{eq:cooling}
\end{equation}
where $T$ is the temperature corresponding to a sound speed of $c_s$ and $\tau_c$ is
constant cooling time, generally written as $\tau_c = \beta \Omega^{-1}$ with $\beta$ a
predefined constant. The sound speed, $c_{so}$, is an effective minimum sound
speed set by the background irradiation and which we generally express in terms of an effective
$Q_{\rm irr}$ where 
\begin{equation}
Q_{\rm irr} = \frac{c_{so} \Omega}{\pi G \Sigma_o}.
\label{eq:Qirr}
\end{equation} 

All the simulations presented here consider a rectangular box of size $L_x = L_y = L = 320$ with
a resolution of $N \times N$ where $N = 1024$, the same as the ``standard" run in \citet{gammie01}. 
The sound speed, or pressure, and surface density are
initially constant and are set such that $Q_{\rm init} = 1$.  The initial perturbations
are introduced through the velocity field which is perturbated from the background
flow by a Gaussian noise component with a subsonic amplitude. Each simulation has
a prescribed cooling time, $\beta$, and an imposed level of external irradiation, $Q_{\rm irr}$.   

\section{Results}
\subsection{Evolution without irradiation}
\begin{figure}
{\epsfig{figure=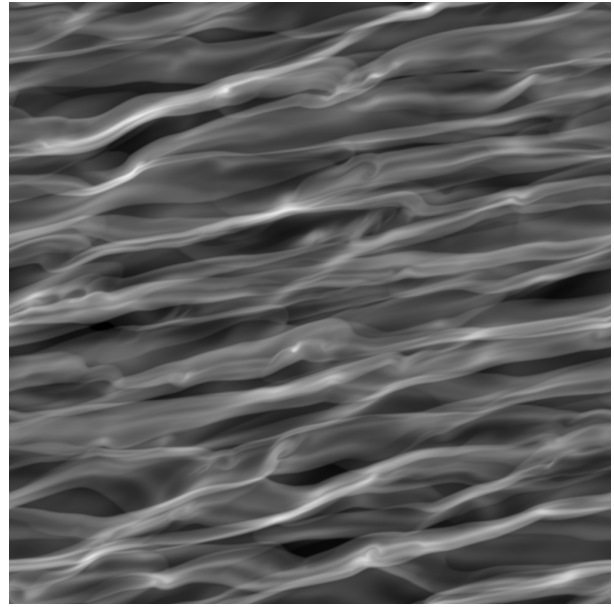, height=80mm}}
\caption{Quasi-steady surface density structure for $\beta = 10$ and $Q_{\rm irr} = 0$. \label{fig:quasi-steady}}
\end{figure}

\begin{figure}
{\psfig{figure=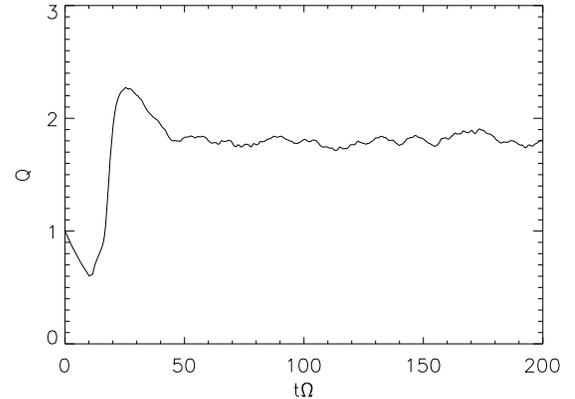, height=60mm}}
\caption{$Q$ profile against time for $\beta = 10$ and $Q_{\rm irr} = 0$. \label{fig:Q_profile}}
\end{figure}

\begin{figure}
{\epsfig{figure=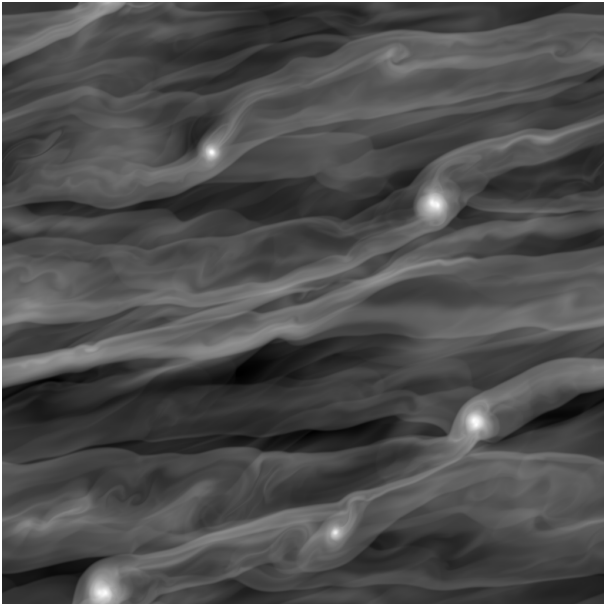, height=80mm}}
\caption{Surface density structure for a simulation, with $\beta = 6$ and $Q_{\rm irr} = 0$, that
has undergone fragmentation. \label{fig:fragment}}
\end{figure}

We consider initially simulations with no external irradiation ($Q_{\rm irr} = 0$) and
vary the cooling time (as measured by $\beta$) to establish the fragmentation boundary
and to investigate the quasi-steady nature of those simulations that don't fragment.
A quasi-steady state is one in which $Q$ settles to
an approximately constant value, generally between $1$ and $2$, and remains at this 
value for many cooling times. In such a state, the instability is active but the system
does not fragment and the instability acts to
transport angular momentum. The disc is regarded as having fragmented
if very dense clumps form, with densities more than 2 orders of magnitude greater than the
average density, and survive for many cooling times.  The presence of a bound clump
also rapidly heats up the surrounding gas and so the $Q$ value in a simulation that
fragments also does not settle to an approximately constant value, but continues to rise.  Fig. {\ref{fig:quasi-steady}}
shows the surface density structure from a simulation that has settled into a quasi-steady state.
In this case $\beta = 10$ and $Q_{\rm irr} = 0$. Fig. \ref{fig:Q_profile} shows the
variation of $Q$ against time illustrating that after an initial burst phase,
$Q$ settles into a quasi-steady state with $Q \sim 1.8$ that persists for many
cooling times. Note that the saturated value of $Q$ is almost a factor 2 larger than what is usually found in 
3D simulations (e.g. Cossins et al 2009), where $Q \sim 1.1$. This is due to the stabilising effect of the finite disc thickness
in 3D simulations \citep{romeo92, mamatsashvili10},  
which dilutes the effect of gravity by $\sim (1+kH)$, where $k$ is the wavenumber of the perturbation. Since for the most unstable 
modes $k \sim 1/H$, we expect a reduction in the linear stability threshold for $Q$ by approximately a factor 2, as observed.  
Fig. \ref{fig:fragment} shows the surface density structure for
a simulation, with $\beta = 6$ and $Q_{\rm irr} = 0$, that has undergone fragmentation.

It can be shown that when a system settles into a quasi-steady state, the shear stress - which we express in terms
of an effective $\alpha$ - satisfies the relationship \citep{pringle81, gammie01}
\begin{equation}
\alpha = \frac{4}{9 \gamma (\gamma - 1) \tau_c \Omega}.
\label{eq:alpha}
\end{equation}
The shear stess can also be determined in each simulation using the Reynolds and gravitational stresses. The average Reynolds
stress is
\begin{equation}
\left<H_{xy}\right> = \left< \Sigma u_x u_y \right>,
\label{eq:Reynolds}
\end{equation}
where $u_x$ and $u_y$ are, again, the velocity perturbations with respect to the background Keplerian
flow.  As described in detail in \citet{gammie01}, the average gravitational shear stress can
be determined in the Fourier domain using
\begin{equation}
\left< G_{xy} \right> = \sum\limits_k \frac{\pi G k_x k_y | \Sigma_k|^2}{|\textbf{\em k}|^3},
\label{eq:grav_stresses}
\end{equation}
where the sum is over all Fourier components. The effective $\alpha$ is then 
\begin{equation}
\alpha = \frac{2}{3 \left< \Sigma c_s^2 \right>}\left(\left<G_{xy}\right> + \left<H_{xy}\right>\right),
\label{eq:alpha_stress}
\end{equation}
which can then be compared with the effective $\alpha$ determined using the imposed cooling time. 

\begin{figure}
{\psfig{figure=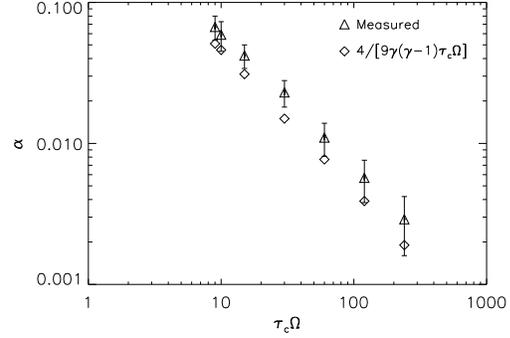, height=50mm}}
\caption{Comparison of the measured effective $\alpha$ values (crosses) 
determined using equation (\ref{eq:alpha_stress}) with that determined from the 
imposed cooling time (diamonds). \label{fig:alpha_tcool}}
\end{figure}

We consider a series of simulations with $\beta$ varying from $4$ to $240$. 
We find that fragmentation occurs for $\beta \le 8$. \citet{gammie01} found fragmentation
for $\beta \le 3$ and the reason for the difference is simply that we have considered a
different specific heat ratio, $\gamma$ \citep{rice05}. For each simulation that does not
fragment we calculate the effective $\alpha$ from the Reynolds and gravitational stresses
using equation (\ref{eq:alpha_stress}) and compare it with that expected from the imposed 
cooling time (e.g., equation (\ref{eq:alpha})).  This is illustrated in Fig. \ref{fig:alpha_tcool} 
in which the measured $\alpha$ values determined using the Reynolds and gravitational stresses
are plotted as triangles, while that expected from the imposed cooling are plotted as diamonds. 
The averaged quantities that are needed to determine the effective $\alpha$ values are written out
by the Pencil Code every 100 iterations, corresponding to a time resolution of $\sim \Omega \Delta t = 0.1$, 
depending on the Courant condition. The measured $\alpha$ values are then determined by averaging over at 
least a few cooling times (generally $5$ or greater, except for the simulations with very long cooling 
times) starting once the system has settled into a quasi-steady state. 

There is a discrepancy between the measured and expected $\alpha$ values, with the mean of the measured
values being higher than that expected. The $1 \sigma$ error bars, however, show that the values are at least
consistent.  The discrepancy is thought to be due to truncation errors.  We have chosen to minimise the
kinematic viscosity and as result some energy is lost at the grid scale \citep{gammie01}.  Figure \ref{fig:alpha_tcool}
does, however, show that the maximum stress that can be attained in a quasi-steady system is $\alpha \sim 0.06$, 
consistent with earlier results \citep{gammie01,rice05}.

\subsection{Evolution of irradiated discs}
In a quasi-steady state in the absence of external irradiation, the imposed cooling is balanced by 
heating due to the Reynolds and gravitational stresses. To consider the influence of 
external irradiation, we impose a combined heating and cooling function of the 
form shown in equation (\ref{eq:cooling}). In a quasi-steady state the following
should then hold \citep{gammie01,mamatsashvili09}
\begin{eqnarray}
\frac{3}{2}\Omega\left<\Sigma u_x u_y + \frac{1}{4 \pi G}\int^\infty_{-\infty} \frac{\partial \phi_x}{\partial x}
\frac{\partial \phi_y}{\partial y} dz \right> = \frac{3}{2}\Omega\left(\left<H_{xy}\right> + \left<G_{xy}\right>\right) \\
=
\frac{\left<\Sigma c_s^2\right> - \left<\Sigma\right> c_{so}^2}{\gamma (\gamma - 1) \tau_c}.
\label{eq:equilibrium}
\end{eqnarray}
Using equation (\ref{eq:alpha_stress}) we can therefore show that, with external irradiation included,
\begin{equation}
\alpha \approx \frac{4}{9 \gamma (\gamma - 1) \Omega \tau_c}\left(1 - \frac{\left<\Sigma\right> c_{so}^2}{\left<\Sigma c_s^2\right>}\right).
\label{eq:alpha_irr}
\end{equation}
If we assume that $\left<\Sigma c_s^2\right> \approx
\left<\Sigma\right> \left<c_s\right>^2$, we can rewrite equation (\ref{eq:alpha_irr}) as
\begin{equation}
\alpha \approx \frac{4}{9 \gamma (\gamma - 1) \Omega \tau_c}\left(1 - \frac{Q_{\rm irr}^2}{Q_{\rm sat}^2}\right),
\label{eq:alpha_Qirr}
\end{equation}
where $Q_{\rm sat}$ is the saturated $Q$ value to which the quasi-steady simulations settle. Since $\Sigma$ and 
$c_s$ are probably correlated, the approximation made to go from Equation (\ref{eq:alpha_irr}) to Equation (\ref{eq:alpha_Qirr})
may not be accurate, but at least gives us an approximate relationship between $\alpha$, $Q_{\rm irr}$ and $Q_{\rm sat}$. The form of 
equation (\ref{eq:alpha_Qirr}) is also consistent with, and a generalisation of, the expression used by \citet{lin90} to represent
viscosity in a self-gravitating disc.  

\begin{figure*}
\begin{center}
\includegraphics[scale = 0.35]{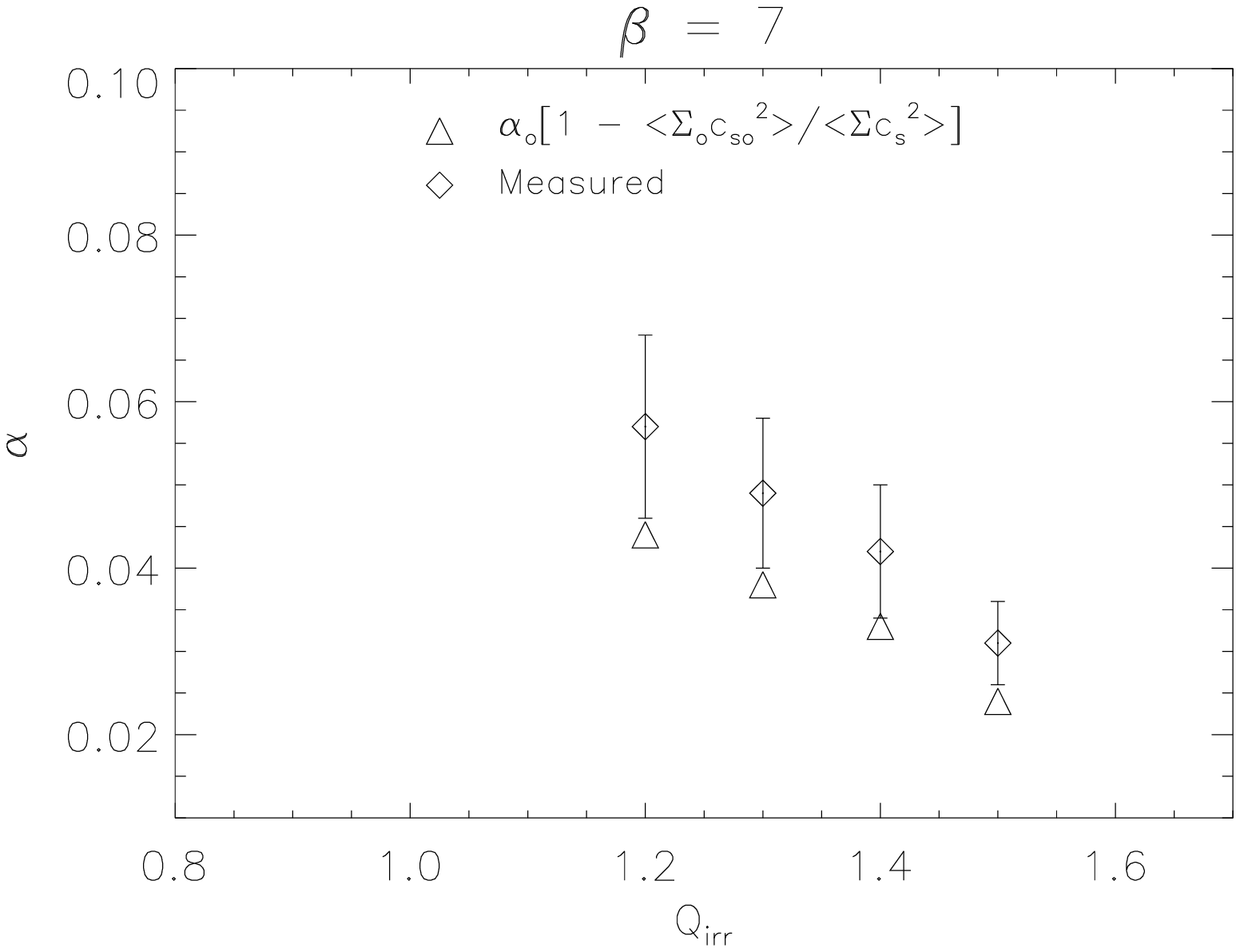}  
\includegraphics[scale = 0.35]{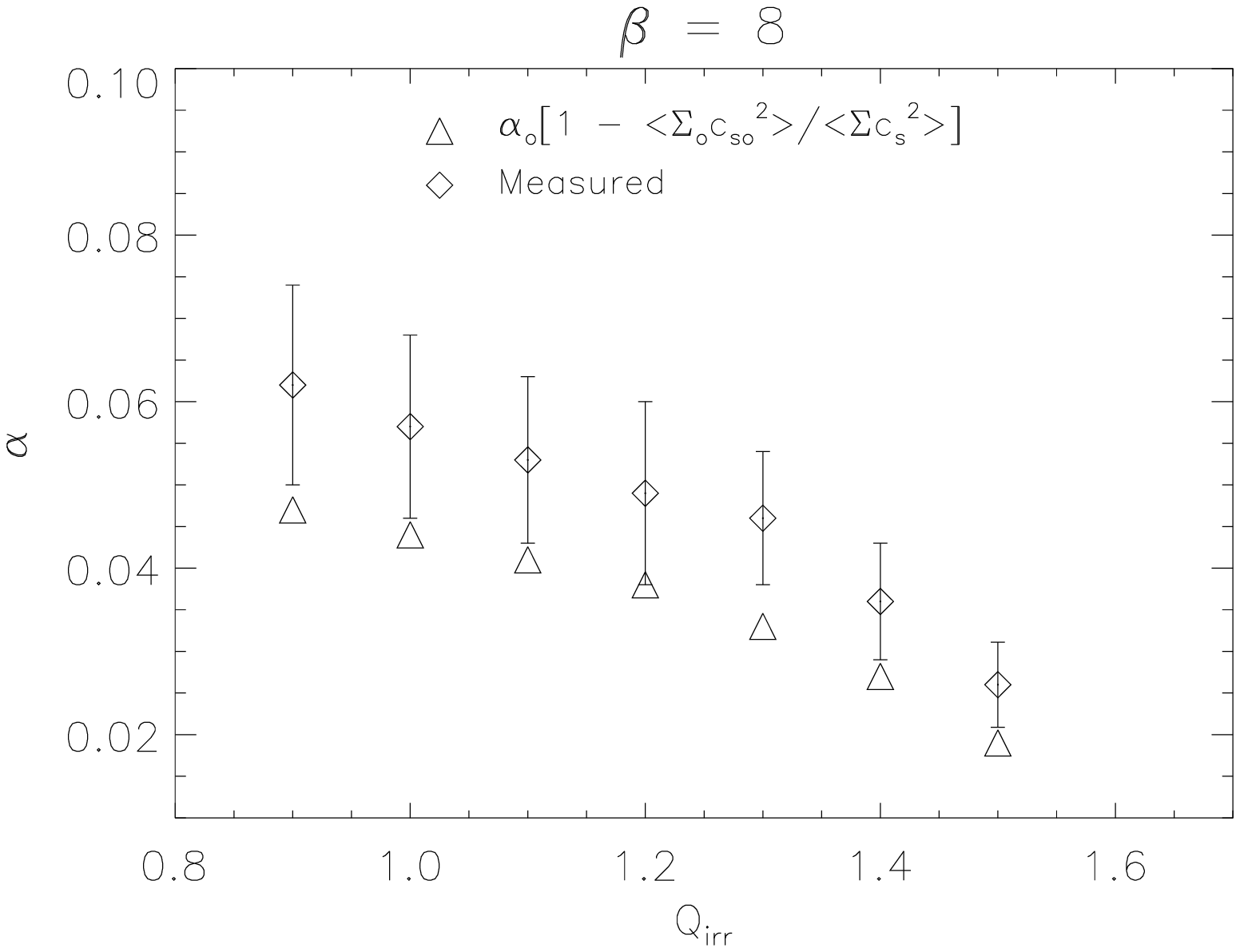} 
\includegraphics[scale = 0.35]{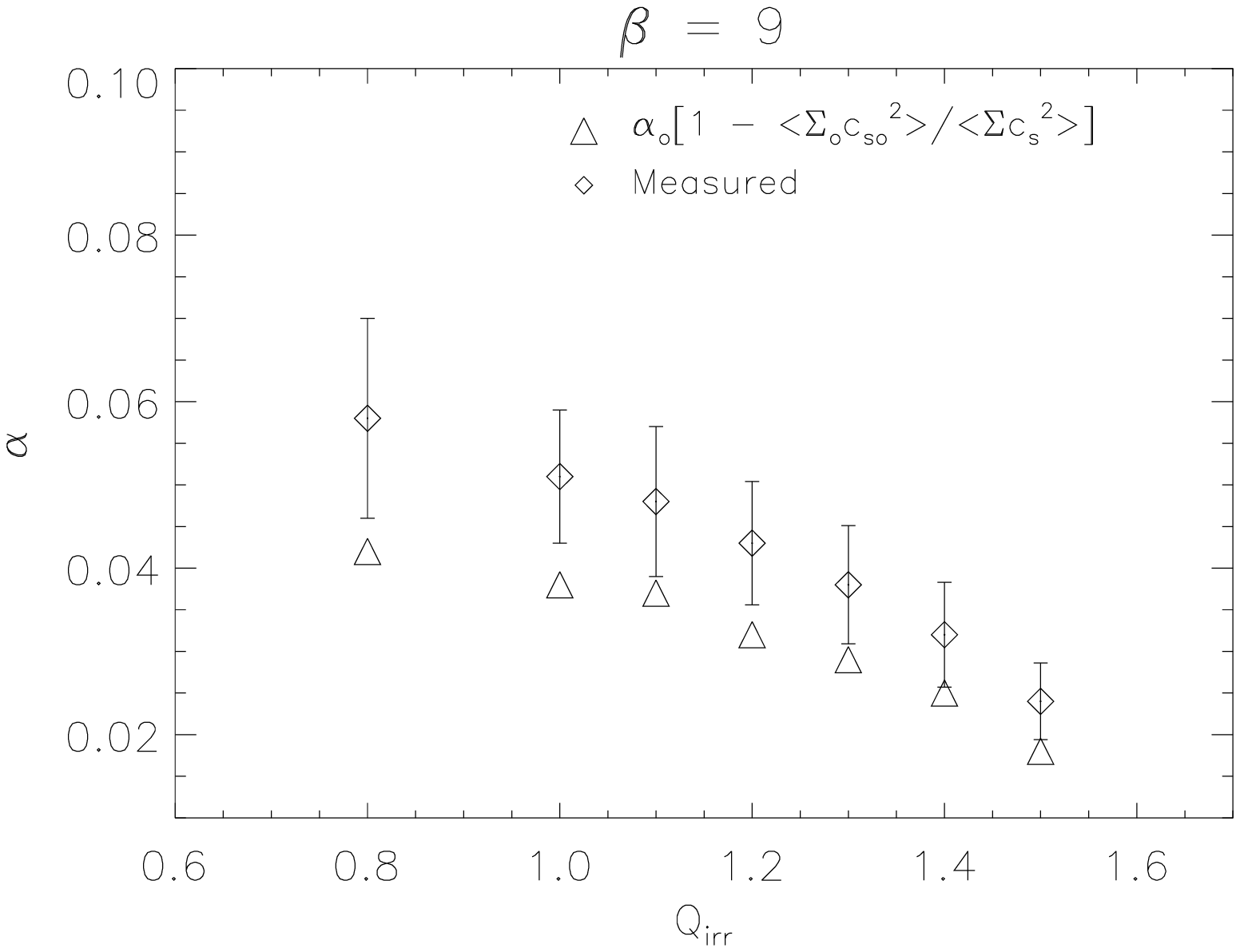} 
\caption{Comparison of measured $\alpha$ values with that determined using the 
imposed cooling time, plotted against $Q_{\rm irr}$ for $\beta = 7$ (top left),
$\beta = 8$ (top right), and $\beta = 9$ (bottom). \label{fig:alpha_tcool_Qirr}}
\end{center}
\end{figure*}

To establish the influence of external irradiation, we consider various values of $\beta$ and for
each value of $\beta$ we consider a number of different values of $Q_{\rm irr}$. Fig. \ref{fig:alpha_tcool_Qirr}
compares the measured $\alpha$ values (determined using the Reynolds and gravitational stresses as
described above) with the expected $\alpha$ values determined using equation (\ref{eq:alpha_irr}),
plotted against $Q_{\rm irr}$ 
for $\beta = 7$ (top left), $\beta = 8$ (top right), and $\beta = 9$ (bottom). 
Again, we only include simulations that settle into a quasi-steady, self-gravitating
state. The diamonds show the mean of the measured values and the error bars are 1 $\sigma$ errors.  The triangles
show the expected value determined using Equation (\ref{eq:alpha_irr}). Once again, the measured values are higher than
those expected but do illustrate, very clearly, that as the level of irradiation increases, the instability weakens -
as expected - and the value of $\alpha$ decreases. A quasi-steady $\alpha$ that exceeds the expected maximum of
$\sim 0.06$ is also never measured.  

\begin{figure}
{\psfig{figure=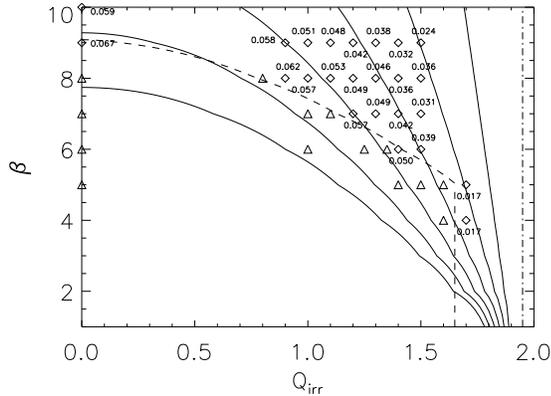, height=60mm}}
\caption{Figure showing contours of constant $\alpha$ determine using equation (\ref{eq:alpha_Qirr})
with $Q_{\rm sat} = 1.9$.  The contour levels are $\alpha = 0.06, 0.05, 0.04, 0.03, 0.02,$ and $0.01$.
The symbols represent all the simulations,
with the diamonds being those that settled into a quasi-steady state and the triangles being
those that fragmented.  The number next to each diamond is the measured $\alpha$ value for that
simulation. The dashed line is the apparent boundary between fragmentation and a quasi-steady,
self-gravitating state in the presence of irradiation. The dash-dot line illustrates the value of
$Q_{\rm irr}$ above which we would expect systems to undergo no instability growth as they would be above
the threshold for linear stability.  \label{fig:contour}}
\end{figure}

Fig.~\ref{fig:contour} summarizes our determination of the fragmentation boundary in the 
$(Q_{\rm irr},\beta)$ plane. 
The plotted contours show lines of constant $\alpha$, determined using equation 
(\ref{eq:alpha_Qirr}) with, using Figure \ref{fig:Q_profile} as a guide, $Q_{\rm sat} = 1.9$.  Ideally we should be using Equation (\ref{eq:alpha_irr}) to set the contours, but
we don't really know an appropriate value for $\left<\Sigma c_s^2\right>$ or how it might vary with $Q_{\rm irr}$, so
Equation (\ref{eq:alpha_Qirr}) provides a suitable approximation. The contour values,
from bottom to top at $Q_{\rm irr} = 0$,  
are $\alpha = 0.06, 0.05, 0.04, 0.03, 0.02$ and $0.01$. The symbols in Fig. \ref{fig:contour} 
show all the simulations that have been performed, with the triangles being those that fragmented and the
diamonds being those that settled into a quasi-steady state.  The number next to each diamond is the
measured $\alpha$ value for that simulation. 

Fig. \ref{fig:contour} illustrates that for every cooling time there is a level of irradiation ($Q_{\rm irr}$)
for which fragmentation does not occur. However, the boundary between fragmentation or a quasi-steady, self-gravitating state does not appear to 
be strictly determined by a maximum value of $\alpha$.  If the boundary was determined largely by a maximum
expected value of $\alpha$, it should lie along one of the solid contours in Fig. \ref{fig:contour}.  
The actual boundary is illustrated by the dashed-line in Fig. \ref{fig:contour}.  As $\beta$ decreases, the
value of $Q_{\rm irr}$ required to halt fragmentation is larger than would be expected.  The maximum
quasi-steady $\alpha$ is also, consequently, smaller than expected. This suggests that the local
cooling rate can influence fragmentation, although it appears that for every cooling time there is
a $Q_{\rm irr}$ that inhibits fragmentation. We would expect that for sufficiently large $Q_{\rm irr}$ 
(e.g., vertical dash-dot line at $Q_{\rm irr} = 1.95$ in Fig. \ref{fig:contour}) there would be no instability
growth as the disc would be maintained
above the threshold for linear instability. Also, for $Q_{\rm irr} = 1.6$, fragmentation
occurs for both $\beta = 4$ and $\beta = 5$, while for $Q_{\rm irr} = 1.7$ a quasi-steady state is achieved
for both values of $\beta$: this suggests that there may be a boundary region (i.e., between $Q_{\rm irr} = 1.7$ and 
$Q_{\rm irr} = 1.95$) in which a quasi-steady, self-gravitating state is achieved for all values of $\beta$. However, it is also
possible that for sufficiently small values of $\beta$ there could be no value of $Q_{\rm irr}$ for which
a quasi-steady (i.e., self-gravitating but non-fragmenting) state is achieved.  If so, the system would either undergo fragmentation or 
be linearly stable. This could have implications for mass loading of systems with very short cooling times.
If there are indeed cooling times for which a quasi-steady state cannot exist then it suggests that such
systems would be unable to viscously respond to increases in surface density and would typically go
from being linearly stable to fragmenting if the surface density increases sufficiently.
The long growth timescales of the instabilty in systems with large values of $Q_{\rm irr}$ is, however,
very long and so we don't attempt to determine here precisely what happens for very short cooling
times.

Ultimately, we have been able to robustly determine the
fragmentation properties of discs with $0 < Q_{\rm irr} \leq 1.6$, reaching values at the top end that are close to the
limit where irradiation entirely quenches gravitational instability. In this range, we find that irradiation
modestly suppresses the critical value of $\beta$ below which fragmentation occurs, from
a value of $\beta_{\rm crit} \simeq 8$ in the absence of irradiation down to a value
$\beta_{\rm crit} \simeq 4$ for the most strongly irradiated disc simulated. The maximum
value of $\alpha$ that a self-gravitating disc can sustain similarly decreases as the
strength of irradiation is increased.

\begin{figure}
{\psfig{figure=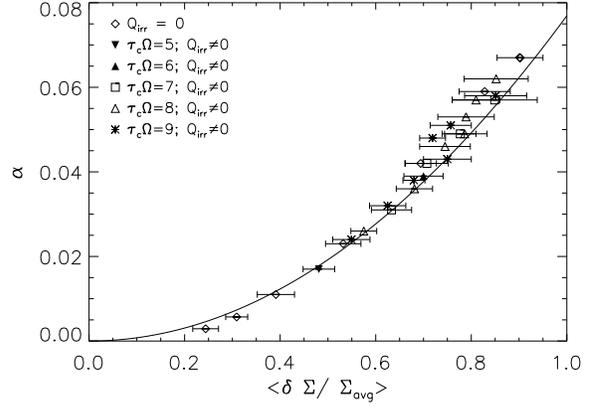, height=60mm}}
\caption{Figure showing the relationship between $\alpha$ and the rms averaged perturbation amplitude,
$\left< \delta \Sigma / \Sigma \right>$. The diamonds show the
values for all the simulations with $Q_{irr} = 0$ while the others symbols represents the
simulations with $Q_{irr} \ne 0$.  The curve is $\alpha \propto \left(\left< \delta \Sigma / \Sigma \right>\right)^2$
illustrating that there is a relationship between the perturbation amplitude and $\alpha$ 
irrespective of whether irradiation is present or not. \label{fig:Sigrms_alpha}}
\end{figure}

\citet{cossins09} suggests that there is a relationship between the perturbation
amplitude ($\delta \Sigma/ \Sigma$) and the cooling parameter, $\beta$. In 
particular they find that the rms averaged perturbation amplitude, 
$\left< \delta \Sigma / \Sigma_{\rm avg} \right>$, obeys the following relationship
\begin{equation}
\left< \frac{\delta \Sigma}{\Sigma_{\rm avg}} \right> \approx \frac{1.0}{\sqrt{\beta}}.
\label{perturb_amp}
\end{equation}
If, as suggested above, the fragmentation boundary is determined primarily by the
effective $\alpha$ rather than by the imposed cooling, $\beta$, we might expect that the 
perturbation amplitude would depend on $\alpha$ rather than $\beta$. Fig. \ref{fig:Sigrms_alpha},
shows $\alpha$ plotted against $\left<\delta \Sigma/\Sigma\right>$ for all the simulations
we performed that did not fragment.  The diamonds are for all the simulations with 
$Q_{\rm irr} = 0$ (i.e., Fig. \ref{fig:alpha_tcool}) while the other symbols are for the
simulations with $Q_{\rm irr} \ne 0$ and correspond to the simulations illustrated in
Fig. \ref{fig:alpha_tcool_Qirr}, although we have also added the $\beta = 4$, $\beta = 5$ and $\beta = 6$ 
simulations that aren't included in Fig. \ref{fig:alpha_tcool_Qirr}.  The values for $\left<\delta \Sigma / \Sigma_{\rm avg} \right>$
were determined by averaging over at least
10 slices starting after the quasi-steady state was reached. The error bars are 1$\sigma$ errors. The curve in Fig. \ref{fig:Sigrms_alpha}
is $\alpha\ \propto \left(\left<\delta \Sigma/\Sigma\right>\right)^2$ and is essentially equivalent
to equation (\ref{perturb_amp}), taken from \citet{cossins09}. Fig. \ref{fig:Sigrms_alpha} 
therefore confirms that the there is a relationship between the perturbation amplitude 
and the effective $\alpha$, rather than the cooling time. That this holds for both
irradiated and non-irradiated discs illustrates that it is universal and is essentially independent
of the condition of thermal equilibrium. 

The reason we have plotted
$\alpha$ against $\left<\delta \Sigma / \Sigma \right>$, rather than the other way around,
is to highlight that the 
fundamental relationship is that the magnitude of the effective $\alpha$
depends only on the magnitude of the density perturbations.  If, as one might expect, fragmentation requires
non-linear density perturbations ($\left< \delta \Sigma / \Sigma \right> \ge 1$) 
this then implies that the maximum $\alpha$ that can be supplied by a quasi-steady system
(i.e., one that does not fragment) is $\alpha \sim 0.06$, consistent with earlier results \citep{gammie01,rice03}.
However, as illustrated above, it appears that for short cooling times fragmentation can occur even if the
expected $\alpha$ is less than $0.06$ which suggests that large, localised perturbations in these systems
can grow to form bound clumps even if the perturbation amplitude in the system as a whole is typically small.  

\begin{figure}
{\psfig{figure=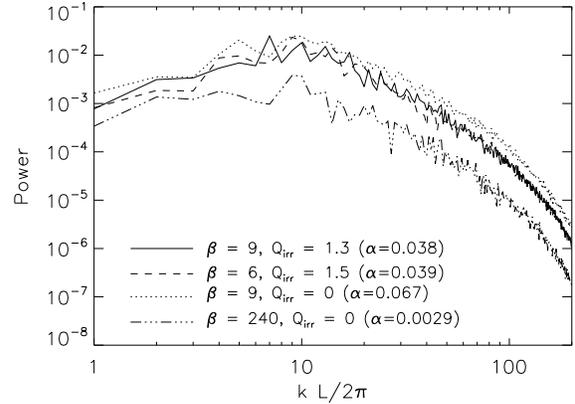, height=60mm}}
\caption{Power spectrum for 4 of the simulations.  The x-axis is in units of $k L/(2 \pi)$ so the peak
at $k L/(2 \pi) \sim 7$ illustrates that most of the shear stress comes from wavelengths significantly smaller
than the model size. The power spectra all have the same basic structure with, as expected, the amplitude increasing with
increasing $\alpha$. The two simulations with approximately the same $\alpha$ values are almost identical illustrating that
the structure of the turbulent state depends primarily on the effective value of $\alpha$.   \label{fig:powerspec}}
\end{figure}

To further illustrate the universality between $\alpha$ and the perturbation amplitudes, we plot, in Fig. \ref{fig:powerspec},
the power spectrum of the perturbations for 4 of the simulations. As in \citet{gammie01} the x-axis is
in units of $k L/(2 \pi)$ and since the power spectra all peak at $k L/(2 \pi) \sim 7$, this illustrates that most of the 
power comes from wavelengths smaller than $L/7$ (i.e., the shear stress come primarily from wavelengths
significantly smaller than the model size).  Fig. \ref{fig:powerspec} also shows that the power spectra all have the same basic form with,
as expected, the amplitude increasing with increasing $\alpha$. Furthermore, the two simulations with the approximately
the same $\alpha$ values (but different values for $\beta$ and $Q_{\rm irr}$) have almost identical power specta. This illustrates
that the structure of the turbulent state does not depend on the cooling rate or the level of irradiation, but is determined
only by the effective value of $\alpha$.  

\section{Conclusions}
For a self-gravitating accretion disc to fragment, gravitational instability must produce strong 
density perturbations that can cool quickly enough to collapse before shear and pressure forces 
disperse them. These requirements for fragmentation can be expressed via two dimensionless 
numbers: a maximum value of $\alpha$, which fixes the amplitude of density fluctuations, and a 
minimum value of $\beta$, which measures the cooling time in terms of the local orbital time scale. 
In the absence of external heating, local thermal equilibrium implies an exact equivalence between 
these descriptions, but this is no longer the case once irradiation becomes significant. In this 
paper, we have used local, two-dimensional simulations of self-gravitating discs to plot the 
fragmentation boundary in both isolated and irradiated discs. We find that the fragmentation 
boundary for irradiated discs does not lie at fixed values of either $\alpha$ or $\beta$. Irradiation 
necessarily weakens the strength of gravitational instability, but whether it stabilizes or destabilizes 
discs against fragmentation depends upon the chosen metric. In terms of the cooling time scale, 
irradiation stabilizes discs: a strongly irradiated disc can avoid fragmentation for cooling times 
almost a factor of two smaller than the limit for a non-irradiated disc. In terms of the maximum 
stress, however, irradiation has the opposite effect: the maximum $\alpha$ sustainable in an 
irradiated disc is lower than that reached in the absence of irradiation. This is somewhat
counterintuitive but is likely a consequence of the universal nature of the form of the power spectrum
of the perturbations.  Although irradiation weakens the instability, it is still possible 
for large, local perturbations to exist. If the cooling time is sufficiently short then 
these large perturbations could collapse to form fragments even if the typical perturbations are
small (i.e., $\alpha_{\rm eff} < 0.06$).

For protoplanetary discs, the implication of our results is to reinforce the conclusion 
that the outer regions of protoplanetary discs are generally unstable to fragmentation 
at the high accretion rates encountered soon after disc formation \citep{clarke09,cossins09b}. 
For a disc forming in a cold environment ($T \approx 10 \ {\rm K}$), the rate of infall 
at radii $r \sim 50 - 100 \ {\rm AU}$ may well exceed the maximum accretion rate 
($\dot M \sim 10^{-7} \ M_\odot \ {\rm yr}^{-1}$) that can be transported inward by the magnetorotational 
instability, even if the MRI is active at these radii despite the effects of ambipolar diffusion \citep{bai11}. 
If so, continued infall will result in the build up of surface density until self-gravity sets in, at 
which point our results imply that irradiation will not save the disc from fragmentation. Irradiation 
will, of course, increase the surface density and temperature of the critically fragmenting disc, 
making it (even) more likely that the outcome of fragmentation will be substellar objects 
rather than massive planets \citep{rice03b,stamatellos07}. Discs forming in substantially 
warmer environments ($T \approx 100 \ {\rm K}$) have a better chance of avoiding 
fragmentation, since the MRI in these systems may be able to transport all the infalling 
gas inward without locally exceeding the threshold surface density for gravitational 
instability. If all other aspects of star formation remained fixed, we would therefore 
expect to see more wide binaries in cool environments, and more young extended 
gas discs in warmer star forming climes.

Finally, we caution that our results strictly apply only to the stability of {\em isolated} 
self-gravitating discs. Protoplanetary discs approaching the threshold surface density 
for gravitational instability do so on account of infall, which need not be steady or 
axisymmetric. The dynamical effects of infall appear to stabilize discs against 
fragmentation \citep{kratter10,harsono11}, at least to some degree, but further 
work is needed to ascertain whether it is possible to avert fragmentation indefinitely 
at radii where isolated local disc models predict collapse.

\section*{Acknowledgments} 
This work was started at the Dynamics of Discs and Planets programme at the 
Isaac Newton Institute for Mathematical Sciences, and we wish to thank both the 
organizers and the Institute for their support. 
P.J.A. acknowledges support from
the NSF (AST-0807471), from NASA's Origins of Solar Systems program
(NNX09AB90G), and from NASA's Astrophysics Theory program (NNX11AE12G).
GRM acknowledges support from the Scottish Universities Physics Alliance (SUPA)
and WKMR from STFC grant ST/H002380/1. This work made use of the facilities of HECToR, the UK's national high-performance computing service, which is provided by UoE HPCx Ltd at the University of Edinburgh, Cray Inc and NAG Ltd, and funded by the Office of Science and Technology through EPSRC's High End Computing Programme. 
The authors would like to thank Anders Johansen for help with the Pencil Code and Charles Gammie, Kaitlin Kratter, and Ruth Murray-Clay for useful disucssions.

\end{document}